\documentclass[pra,twocolumn,supersciptaddress]{revtex4}
\usepackage{graphicx}

\newcommand{\ket}[1]{ \left. | #1 \right\rangle }

\begin{document}

\title{A Proposal for an Optical Implementation of an Universal Quantum Phase Gate}

\author{S. Rebi\'{c}, D. Vitali, C. Ottaviani, P. Tombesi}
\address{INFM and Dipartimento di Fisica, Universit\`{a} di Camerino, I-62032 Camerino, Italy,\\ E-mail: stojan.rebic@unicam.it}

\author{M. Artoni, F. Cataliotti}
\address{European Laboratory for Non-linear Spectroscopy, via N. Carrara, I-50019 Sesto Fiorentino, Italy}

\author{R. Corbal\'{a}n}
\address{Departament de F\'{i}sica, Universitat Aut\`{o}noma de Barcelona, E-08193, Bellaterra, Spain}

\begin{abstract}
Large optical nonlinearities occurring in a coherently prepared atomic system are shown to produce phase shifts of order $\pi$. Such an effect may be observed in ultracold rubidium atoms where it could be feasibly exploited toward the realization of a polarization phase gate. 
\end{abstract}

\maketitle

\section{Introduction \label{sec:intro}}

A great effort has recently gone into the search for practical architecture for quantum information processing systems. In this paper we focus on optical implementations of quantum information processing systems -- in particular quantum phase gate\cite{NielsenChuang}. One of the possible ways to realize this system requires strong interaction of the photonic qubits. Sufficiently strong interactions have been unavailable until recently. The effect of electromagnetically induced transparency (EIT)\cite{Harris97} and its use in the implementation of nonlinear optical interactions opened a way for generation of large optical nonlinearities, and hence strong photon-photon interactions~\cite{Schmidt96}.

A significant cross-phase modulation is the key ingredient for the implementation of a quantum phase gate between two optical qubits. Such a cross-phase modulation could be realized exploiting the cross-Kerr effect whereby an optical field acquires a phase shift conditioned to the state of another optical field. The relevant gate transformation is defined through the following input-output relations $|i\rangle _{1}|j\rangle _{2}
\rightarrow \exp\left\{i \phi_{ij} \right\}|i\rangle _{1}|j\rangle _{2}$, where $i,j=0,1$ denote the qubit basis. In particular, this becomes a universal two-qubit gate, that is a gate able to entangle two initially factorized qubits, when the conditional phase shift $\phi=\phi_{11}+\phi_{00}-\phi_{10}-\phi_{01}$ becomes different from zero\cite{NielsenChuang,Lloyd95}.

\section{Optical Quantum Phase Gate \label{sec:qpg}}

\begin{figure}[t!]
\includegraphics[scale=0.7]{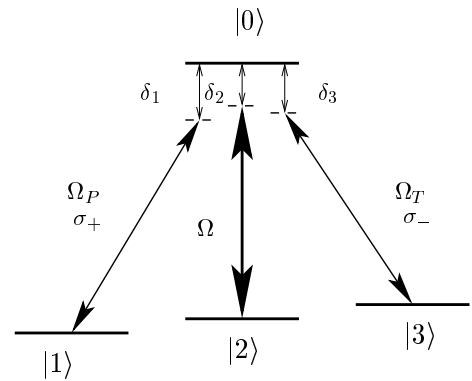}
\caption{Energy level scheme for a tripod. Detunings $\delta_j= \omega_0-\omega_j-\omega_j^{(L)}$ denote the laser (frequency $\omega_j^{(L)}$) detunings from the respective transitions $|j\rangle \leftrightarrow |0\rangle$. $\Omega$'s denote  Rabi frequencies of the respective fields. Cross-phase modulation is achieved between the {\em probe} and the {\em trigger} fields (driving transitions $\ket{1}\rightarrow\ket{0}$ and $\ket{3}\rightarrow\ket{0}$, respectively) \label{fig:tripod} }
\end{figure}

A natural choice for encoding binary information in optical beams consists in using the polarization degree of freedom, in which case the two logical basis states $|0\rangle $ and $|1\rangle$ of the above gate transformation correspond to two orthogonal light polarizations. A possible experimental implementation can be realized with the tripod scheme shown in Fig.~\ref{fig:tripod} by using $^{87}$Rb atoms confined in a temporal dark SPOT (Spontaneous-force Optical Trap). This is a magneto-optical trap (MOT) where the repumping beam has been temporarily shut off\cite{Ketterle93}. In such a trap cold atoms are transferred in the $\ket{5S_{1/2}, F = 1, m = \{ -1, 0, 1 \}}$ state(s) of $^{87}$Rb while density is increased with respect to a conventional MOT. In this case states $|1\rangle$, $|2\rangle$ and $|3\rangle$ correspond to the ground state Zeeman sublevels $|5S_{1/2}, F = 1, m = \{ -1, 0, 1 \} \rangle$, and state $|0\rangle$ corresponds to the excited state $|5P_{3/2}, F = 0 \rangle$. The atoms are available for just a few milliseconds which is long compared with  the typical microsecond time scales involved in our proposed experiment.

A universal QPG could be implemented when a significant and non-trivial cross-phase modulation between probe and trigger fields arises but only for one of the four input probe and trigger polarization configurations. This occurs for our tripod configuration of Fig.~\ref{fig:tripod} only when the probe is $\sigma^+$ polarized and the trigger is $\sigma^-$ polarized. When the probe has instead a $\sigma^-$ polarization [Eq.~(\ref{eq:sigmamm})], that is to say the ``wrong'' polarization, there is no sufficiently close level it may couple to and hence the corresponding pulse will acquire the trivial (vacuum) phase shift $\phi_{0}^{P}=k_{P}l$, where $l$ is the length
of the medium. The trigger pulse with the ``correct'' $\sigma^-$ polarization, on the other hand, will acquire in this case the linear phase shift
\begin{equation}
\label{eq:linphase}
    \phi_{lin}^T = k_Tl \left( 1 + 2\pi \chi_T^{(1)} \right).
\end{equation}
It is worth noticing here that for sufficiently narrow probe and trigger laser linewidths and nearly equal detunings as used in our scheme, cross-phase modulation between the two $\sigma^-$ polarized probe and trigger pulses [Eq.~(\ref{eq:sigmamm})] does not occur for sublevels Zeeman shifts larger than (half) the EIT transparency bandwidth. Owing to the fact that such a bandwidth is typically smaller than $\gamma$, or even much smaller as in the case that we examine ($\sim 0.1\gamma$), cross-Kerr nonlinearities for the case of a wrong probe polarization [Eq.~(\ref{eq:sigmamm})] can readily be avoided for sufficiently large Zeeman splittings\footnote{Right and wrong polarizations are distinguished by their frequencies so even when a $\sigma^-$ polarized probe, e.g., couples to the $\ket{3} \rightarrow \ket{0}$ transition, it would fall outside the transparency windows already for Zeeman splittings of several $\gamma$'s leading to a vanishing cross-Kerr modulation.}. This is realistic assumption, given that the Zeeman splitting of levels $\ket{1}$ and $\ket{3}$ is typically as large as $20\gamma$, thus giving a two orders of magnitude difference between the size of the transparency window and the frequency of the (``wrong") trigger transition. Zeeman shift of this size also insures that the ``wrong" polarization qubit is not absorbed outside the transparency window. The case of a wrong $\sigma^+$ polarized trigger [Eq.~(\ref{eq:sigmapp})] can be discussed in just the same way leading to a vacuum shift $\phi_{0}^{T}$ and to a linear shift $\phi_{lin}^P$ which obtain from the $\phi_{0}^{P}$ and $\phi_{lin}^T$ above upon interchanging $P \leftrightarrow T$. When probe and trigger both have the ``wrong'' polarization, i.e. probe is $\sigma^{-}$ polarized and trigger $\sigma^{+}$ polarized, there is no sufficiently close level to which probe and trigger can be coupled to and the fields acquire the trivial vacuum phase shift $\phi_{0}^{j}=k_{j}l$, $j=P,T$.

A probe and a trigger polarized single photon wave packets form a qubit\cite{Ottaviani03}
\begin{equation}
\ket{\psi_i} = \alpha_i^+ \ket{\sigma^+}_i + \alpha_i^- \ket{\sigma^-}_i, \ \ \ \ \ \ i=\{P,T\}.
\end{equation}
This qubit is a superposition of two circularly polarized states
\begin{equation}
\ket{\sigma^\pm}_i = \int d\omega \, \xi_i (\omega) a_\pm^\dagger(\omega) \ket{0},
\end{equation}
where $\xi_i (\omega)$ is a Gaussian frequency distribution of incident wave packets, centered at frequency $\omega_i$. Traversing the atomic medium of length $l$, the photon field operator undergoes a transformation
\begin{equation}
a_\pm(\omega) \rightarrow a_\pm(\omega) \exp{\left\{ i\frac{\omega}{c}\int_0^l dz\, n_\pm(\omega,z) \right\}}.
\end{equation}
The real part of the refractive index $n_\pm$ can be assumed to vary slowly over the bandwidth of the wave packets $n_\pm(\omega,z) \approx n_\pm(\omega_i,z)$, giving rise to a phase shift on a circularly polarized states $\ket{\sigma^\pm}_i \rightarrow e^{-i\phi^i_\pm}\ket{\sigma^\pm}_i$, where 
\begin{equation}
\phi^i_\pm =\frac{\omega}{c}\int_0^l dz\, n_\pm(\omega_i,z).
\end{equation}
For a Gaussian trigger pulse of time duration $\tau_T$ and Rabi frequency $\Omega_T$, moving with group velocity $v_g^T$, the nonlinear probe phase shift can be written as
\begin{equation}
    \phi_{nlin}^P = k_Pl
\frac{\pi^{3/2}\hbar^2|\Omega_T|^2}{4|\bm{\mu}_T|^2}\,
\frac{\rm{erf}[\zeta_P]}{\zeta_P} \, {\rm Re}[\chi_P^{(3)}],
\end{equation}
where $\zeta_P = (1-v_g^P/v_g^T)\sqrt{2}l/v_g^P\tau_T$. The trigger shift is obtained upon interchanging $P \leftrightarrow T$ in the equation above, namely
\begin{equation}
    \phi_{nlin}^T = k_Tl
\frac{\pi^{3/2}\hbar^2|\Omega_P|^2}{4|\bm{\mu}_P|^2}\,
\frac{\rm{erf}[\zeta_T]}{\zeta_T} \, {\rm Re}[\chi_T^{(3)}],
\end{equation}
where also the expression for $\zeta_T$ has to be changed accordingly. The expressions for nonlinear susceptibilities are\cite{Rebic03}
\begin{eqnarray}
    \chi_P^{(3)} &=& \mathcal{N}\,
\frac{|\bm{\mu}_P|^2|\bm{\mu}_T|^2}{\hbar^3\epsilon_0} \times
\frac{1}{2}\frac{\Delta_{12}/\Delta_{13}}{\Delta_{10}\Delta_{12}-|\Omega|^2}
\nonumber \\
&\ & \times \left(
\frac{\Delta_{12}}{\Delta_{10}\Delta_{12}-|\Omega|^2}+\frac{\Delta_{23}}{\Delta_{30}^*\Delta_{23}-|\Omega|^2}
\right), \label{eq:kerrsuscp} \\
    \chi_T^{(3)} &=& \mathcal{N}\,
\frac{|\bm{\mu}_T|^2|\bm{\mu}_P|^2}{\hbar^3\epsilon_0} \times
\frac{1}{2}\frac{\Delta_{23}^*/\Delta_{13}^*}{\Delta_{30}\Delta_{23}^*-|\Omega|^2}
\nonumber \\
&\ &\times \left(
\frac{\Delta_{12}^*}{\Delta_{10}^*\Delta_{12}^*-|\Omega|^2}+\frac{\Delta_{23}^*}{\Delta_{30}\Delta_{23}^*-|\Omega|^2}
\right) , \label{eq:kerrsusct}
\end{eqnarray}
where $\mathcal{N}$ is density of a medium, $\bm{\mu}_{P,T}$ are electric dipole matrix elements for probe and trigger transitions while the complex detunings are defined as $\Delta_{j0} = \delta_j +i\gamma_{j0}$ and $\Delta_{kj} = \delta_j - \delta_k - i\gamma_{kj}$ for $k,j = 1,2,3$. Also, $\gamma_{j0}$ are spontaneous emission rates and $\gamma_{kj}$ are dephasing rates.

The truth table for a polarization QPG that uses our tripod configuration reads as
\begin{eqnarray}
    |\sigma^-\rangle_P |\sigma^-\rangle_T &\rightarrow&
e^{-i(\phi_0^P+\phi_{lin}^T)}|\sigma^-\rangle_P |\sigma^-\rangle_T, \label{eq:sigmamm}\\
    |\sigma^-\rangle_P |\sigma^+\rangle_T &\rightarrow&
e^{-i(\phi_0^P+\phi_0^T)}|\sigma^-\rangle_P |\sigma^+\rangle_T, \label{eq:sigmamp}\\
    |\sigma^+\rangle_P |\sigma^+\rangle_T &\rightarrow&
e^{-i(\phi_{lin}^P+\phi_0^T)}|\sigma^+\rangle_P |\sigma^+\rangle_T, \label{eq:sigmapp}\\
    |\sigma^+\rangle_P |\sigma^-\rangle_T &\rightarrow&
e^{-i(\phi_+^P+\phi_-^T)}|\sigma^+\rangle_P |\sigma^-\rangle_T, \label{eq:sigmapm}
\end{eqnarray}
with $\phi_+^P = \phi_{lin}^P + \phi_{nlin}^P$ and $\phi_-^T = \phi_{lin}^T + \phi_{nlin}^T$ and where the conditional phase shift is given by
\begin{equation}
\phi = \phi_+^P + \phi_-^T - \phi_{lin}^P - \phi_{lin}^T.
\end{equation}
Notice that only the nonlinear shifts contribute to $\phi$. The truth table of Eqs.~(\ref{eq:sigmamm}-~\ref{eq:sigmapm}) differs from that of Ottaviani {\it et al.}\cite{Ottaviani03} only for the presence of a linear phase shift for the trigger, arising from the fact that also level $\ket{3}$ is populated with one half of the atoms.

In the $^{87}$Rb level configuration chosen above, the decay rates are equal $\gamma_{j0}= \gamma$ and we take for simplicity equal and small dephasing rates $\gamma_{ij}\simeq \gamma_d = 10^{-2} \gamma$. For $\Omega_P \approx \Omega_T = 0.1 \gamma$, $\Omega = \gamma$, and detunings $\delta_1 = 20.01\gamma$, $\delta_2 = 20\gamma$, $\delta_3 = 20.02\gamma$ we obtain a conditional phase shift of $\pi$ radians over an interaction length $l = 1.6$ mm at a density $\mathcal{N} = 3 \times 10^{13}$ cm$^{-3}$. With these parameters, group velocities are essentially the same, giving $\rm{erf}[\zeta_P]/\zeta_P = \rm{erf}[\zeta_T]/\zeta_T \approx 2/\sqrt{\pi}$. This choice of parameters corresponds to the case where probe and trigger have a mean amplitude of about one photon when the beams are tightly focused ($\sim$1 $\mu$m) and with a time duration in the microsecond scale. 

In addition, it is worthwhile noting that a classical phase gate could be implemented by using more intense probe and trigger pulses. For Rabi frequencies $\Omega_P \approx \Omega_T = \gamma$, $\Omega = 4.5 \gamma$, and detunings $\delta_1 = 10.01\gamma$, $\delta_2 = 10\gamma$, $\delta_3 = 10.02\gamma$, a conditional phase shift of $\pi$ radians, over the interaction length $l = 0.7$ cm, density $\mathcal{N} = 3 \times 10^{12}$ cm$^{-3}$ is obtained. Again, with these parameters, group velocities are the same. 

Both sets of parameters could be realized with cold atoms in a temporal dark SPOT of a MOT. Alternatively, a gas cell of standard length between 2.5 cm and 10 cm can be considered, but the increase in length is then compensated with a lower density. In this case one has to take care to use all co-propagating laser beams to cancel the first order Doppler effect\cite{Arimondo96}. This shows that a demonstration of a deterministic polarization QPG can be made using present technologies.

\section{Conclusion}

In this paper we have presented the proposal for a realization of an optical polarization phase gates. The requirements for a gate operation have been discussed in detail. The gate can be realized using present technologies in a cold atomic sample of $\ ^{87}$Rb atoms.

\section*{Acknowledgements}
We acknowledge enlightening discussions with P. Grangier, F. T. Arecchi and M. Inguscio. We greatly acknowledge support from the MURST (\emph{Actione Integrada Italia-Spagna}), the MIUR (PRIN 2001 \emph{Quantum Communications with Slow Light}) and by MCyT and FEDER (\emph{project BFM2002-04369-C04-02}).


\begin{thebibliography}{0}
\bibitem{NielsenChuang}M.~A. Nielsen and I.~L. Chuang,
{\it Quantum Computation and Quantum Information}, (Cambridge University Press, Cambridge, 2000.)

\bibitem{Harris97}S. Harris,
Physics Today {\bf 50}(7), 36 (1997).

\bibitem{Schmidt96}H. Schmidt and A. Imamo\u{g}lu, Opt. Lett. {\bf 21}, 1936 (1996); 
 H.~Wang, D.~Goorskey and M.~Xiao, Phys. Rev. Lett.  {\bf 87}, 073601 (2001); 
 H. Kang and Y. Zhu, Phys. Rev. Lett. {\bf 91}, 093601 (2003).

\bibitem{Lloyd95}S. Lloyd, Phys. Rev. Lett. {\bf 75}, 346 (1995).
Q.~A. Turchette, C.~J. Hood, W. Lange, H. Mabuchi and H.~J. Kimble, Phys. Rev. Lett. {\bf 75}, 4710 (1995).

\bibitem{Ketterle93}W. Ketterle, K.~B. David, M.~A. Joffe, A. Martin and D. Pritchard,
Phys. Rev. Lett. {\bf 70}, 2253 (1993).

\bibitem{Ottaviani03} C.~Ottaviani, D.~Vitali, M.~Artoni, F.~Cataliotti and P.~Tombesi,
Phys. Rev. Lett. {\bf 90}, 197902 (2003).

\bibitem{Rebic03} S.~Rebi\'{c}, D.~Vitali, C.~Ottaviani, P.~Tombesi, M.~Artoni, F.~Cataliotti and R. Corbal\`{a}n, ``Polarization phase gate with a tripod atomic system'', quant-ph/0310148.

\bibitem{Arimondo96} E. Arimondo, in {\it Progress in Optics XXXV},
 eds.~E. Wolf and L. Mandel, (Elsevier Science, Amsterdam, 1996), pp. 257-354.

\end{thebibliography}
\end{document}